\documentclass[conference]{IEEEtran}
\IEEEoverridecommandlockouts
\usepackage{float}
\usepackage{cite}
\usepackage{amsmath,amssymb,amsfonts}
\usepackage{algorithmic}
\usepackage{graphicx}
\usepackage{textcomp}
\usepackage{xcolor}

\usepackage{booktabs}
\usepackage{multirow}

\def\BibTeX{{\rm B\kern-.05em{\sc i\kern-.025em b}\kern-.08em
    T\kern-.1667em\lower.7ex\hbox{E}\kern-.125emX}}
\begin{document}


\title{Sensing Multi-modal Mobility Patterns: A Case Study of Helsinki using Bluetooth Beacons and a Mobile Application}

\author{\IEEEauthorblockN{1\textsuperscript{st} Zhiren Huang}
\IEEEauthorblockA{\textit{Dept. of Computer Science} \\
\textit{Aalto University}\\
Espoo, Finland \\
0000-0002-7868-1630}
\and
\IEEEauthorblockN{2\textsuperscript{nd} Alonso Espinosa Mireles de Villafranca}
\IEEEauthorblockA{\textit{Dept. of Built Environment} \\
\textit{Aalto University}\\
Espoo, Finland \\
0000-0002-4521-0234}
\and
\IEEEauthorblockN{3\textsuperscript{rd} Charalampos Sipetas}
\IEEEauthorblockA{\textit{Dept. of Built Environment } \\
\textit{Aalto University}\\
Espoo, Finland \\
0000-0002-9829-3483}
}

\maketitle

\begin{abstract}

Detailed understanding of multi-modal mobility patterns within urban areas is crucial for public infrastructure planning, transportation management, and designing public transport (PT) services centred on users' needs.
Yet, even with the rise of ubiquitous computing, sensing urban mobility patterns in a timely fashion remains a challenge. Traditional data sources fail to fully capture door-to-door trajectories and rely on a set of models and assumptions to fill their gaps. This study focuses on a new type of data source that is collected through the mobile ticketing app of HSL, the local PT operator of the Helsinki capital region. HSL's dataset called TravelSense, records anonymized travelers' movements within the Helsinki region by means of Bluetooth beacons, mobile phone GPS, and phone OS activity detection. 
In this study, TravelSense dataset is processed and analyzed to reveal spatio-temporal mobility patterns as part of investigating its potentials in mobility sensing efforts. The representativeness of the dataset is validated with two external data sources - mobile phone trip data (for demand patterns) and travel survey data (for modal share). Finally, practical perspectives that this dataset can yield are presented through a preliminary analysis of PT transfers in multimodal trips within the study area.


\end{abstract}


\section{Introduction}
This study focuses on a new app-based dataset (called "TravelSense") that is derived from a mobile ticketing app launched and operated by HSL, the local public transport (PT) operator in Helsinki, Finland. The Helsinki PT network primarily uses tap-in-only smartcards, which leads to challenges in reconstructing origin-destination (OD) matrices and demand patterns when using automated fare card (AFC) data only. For example, missing parts of a traveler's full trajectory and can be captured by the mobile application. The app is characterized by a user-centric design that allows the operator to complement available datasets with complete journey data collected from users who have agreed to share this information. 

The goals of this study are to a) present this new dataset, b) describe methods and assumptions needed for processing and analyzing it properly, c) utilize it for simple transport applications in order to compare it with results obtained from alternative well-established sources, and d) offer preliminary insights on its potential usage in critical transport research areas. The focus is on constructing exact traveler trajectories which can be further processed for quantifying transfers at PT stop and hub level. Transfers are the core of multi-modal trips and are often a hindrance, associated with high disutilities for passengers when choosing PT instead of less sustainable modes. 

Contributions of this study are a) the analysis of an emerging data source with an emphasis on issues and challenges associated with it, and b) the investigation of its capability for filling existing gaps of traditional and widely used datasets within PT systems. Such efforts are critical for better understanding the PT user patterns, behaviors and needs in order to improve PT performance through targeted approaches referring to planning, design and operation of PT services.

\section{Literature review}

With the rapid development of information and communications technology (ICT) infrastructure, a wide range of technologies are applied to capture the timely mobility patterns within urban areas. According to detailed literature reviews \cite{Zheng2014, Zheng2016, Welch2019} common data sources for sensing urban mobility (across different transport modes) include smartcard data, sensor data (e.g., Bluetooth and WiFi), and mobile phone data (e.g., call detail records and signaling). Each data source has drawbacks, with some of them only capturing a single mode and others lacking detailed mode information even when capturing multimodal aspects.
In the field of PT, a common approach is for transport authorities to install automated fare control (AFC) system where passengers use smartcards. Uses of smartcard data include estimation of route choices \cite{Zhao2017}, classification of passenger behavior and its changes over time \cite{Briand2017}, and passenger responses to PT disruptions such as delays \cite{Tian2018}. However, studies such as these use data that are limited to within the PT system, with information before entering or after exiting PT systems missing. Moreover, those data sources are usually transaction-based, hence, only record information when passenger taped their smartcards or mobile phones. A common problem is the uncertainty of destination stops, especially in systems where tap-out is not required \cite{Zheng2018, Tu2019}.


To complement the sparseness of smartcard data, Bluetooth beacons and WiFi AP can be used to continuously sense users' mobile phones. Ellersiek et al. \cite{Ellersiek2013} performed a field study in a zoo using Bluetooth sensors where they analyze visitors' trajectories. Kostakos et al. \cite{kostakos2013} trialed a Bluetooth transceiver system on-board buses for sensing passengers, and while they show some promising results in OD matrix reconstruction, they found that only around 12\% of passengers had Bluetooth enabled. More recently, Tu et al. \cite{Tu2019} used buses' WiFi to identify passengers' boarding stops and infer partial destinations. However, those devices are also usually installed only on specific stops (e.g., metro stations) or transit vehicles (e.g., buses).


Unlike smartcard or Bluetooth/WiFi data, which only captures data within specific systems or sites, mobile phone data are considered to be more representative and are widely used in OD \cite{Wang2012, Iqbal2014, Alexander2015, Jiang2017, Huang2018, Wang2018} and population estimation \cite{Deville2014, Bergroth2022}. Mobile phone call detail records (CDR) can record the served cell tower location when a user calls, messages, or accesses the internet. Previous studies \cite{Wang2012, Iqbal2014, Alexander2015} use CDR for OD estimation and activity-based modeling \cite{Jiang2017}. Mobile phone signaling data can record tower switch information as opposed to CDR, which can yield more accurate trip information \cite{Huang2018, Wang2018}.

Although OD estimation using mobile phone data is in a mature stage, transportation mode inference remains challenging. This is partly due to cell tower locations only approximating user locations as well as the sparcity of CDR trajectories (in comparison to GPS, for example). Efforts to address this include discrete choice modelling \cite{Qu2015} using CDR and transport network data to identify transportation modes (driving, PT, and walking), and Bayesian inference of road and rail OD matrices \cite{Bachir2019}.
To overcome some drawbacks of CDR data, some studies focused on mining GPS trajectories from different mobile apps like travel experience \cite{Zheng2008a, Zheng2008b, Huang2020} and navigation applications \cite{Wang2022}.
Using feature engineering and machine learning approaches \cite{Zheng2008a, Zheng2008b, Huang2020, Wang2022}, transport modes of mobile phone GPS trajectories can be identified with high accuracy. However, GPS signals are vulnerable to loss in complex environments, such as underground when users take the subway \cite{MSA2011}.

Given that each data source has its shortcomings, some studies explore how to combine data sources to capture more comprehensive mobility patterns. Huang et al. \cite{Huang2018} fused mobile phone signaling data, smartcard data, and taxi GPS to model whole population-level human mobility within a mega city. Versichele et al. \cite{Versichele2014} combined Bluetooth data and WiFi data to enhance the accuracy of pedestrian flow estimation. Sipetas et al. \cite{Sipetas2020} fused train operations data with data from object detection tools to address crowding phenomena. Meng et al. \cite{Meng2022} used travel survey trajectories, geo-tagged Twitter data, and Point of Interest (POI) data to infer the trip purposes. Poonwala et al. \cite{Poonwala2016} fused AFC and CDR to reconstruct journeys and calibrate route choice models. Hence, mining the potential of each data source and combining different data sources can contribute to more comprehensive mobility patterns.

\section{Data collection}

\subsection{Study site description}
\label{sec:site}


\begin{figure}
    \centering
    \includegraphics[width=8cm]{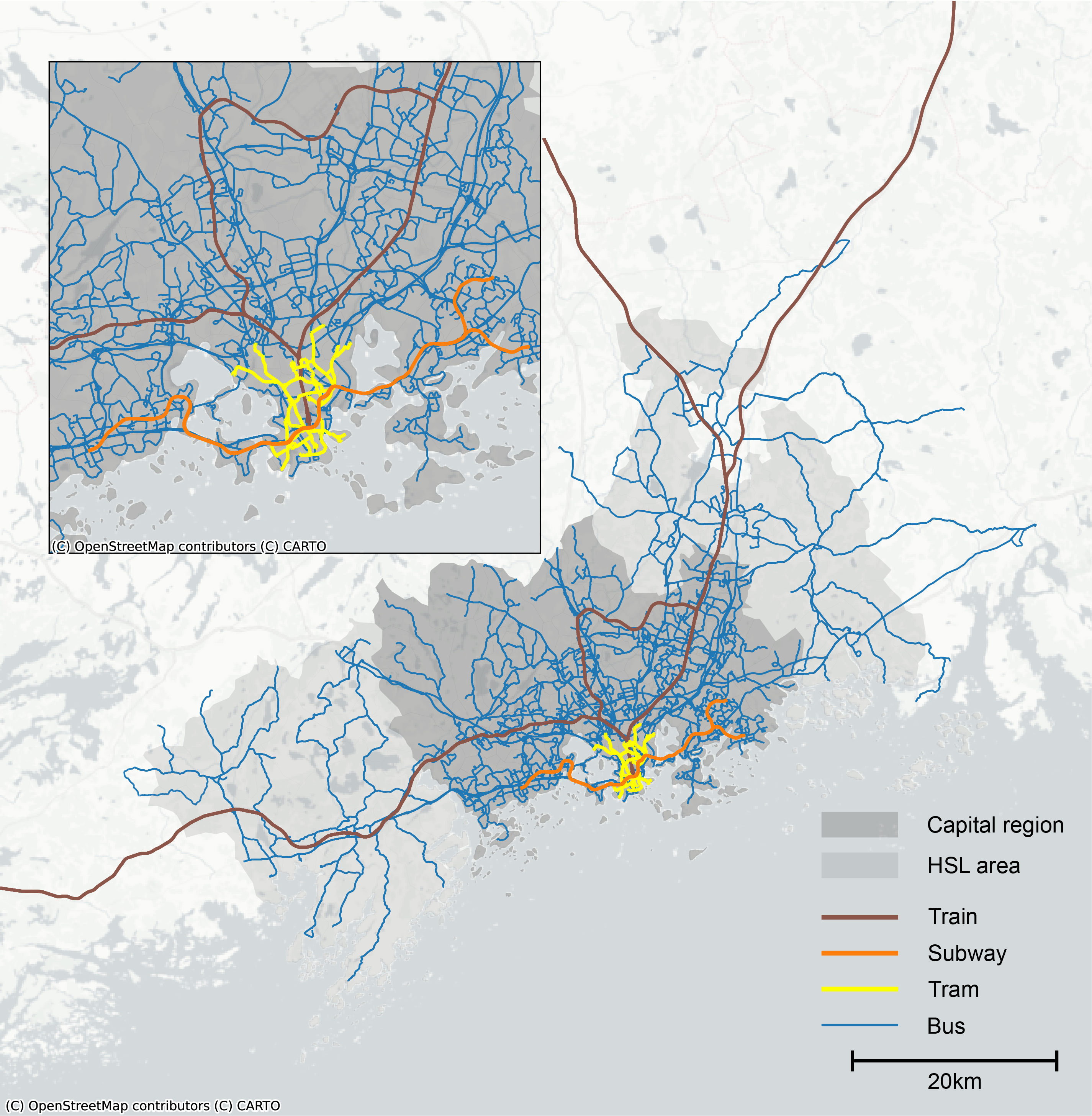}
    \caption{The study area and the PT network including different transportation modes: train (brown), subway (orange), bus (blue), and tram (yellow).}
    \label{fig:study_area}
\end{figure}

The capital region of Finland is a metropolitan area, which is composed of 4 municipalities (i.e., cities): Helsinki, the capital city, as well as Espoo, Vantaa, and Kauniainen. The land area equals approximately 770 km$^2$ with a population of approximately 1.2 million inhabitants. The local mobility services include fixed PT (metro, tram, train, bus, and ferry), micro-mobility (shared e-scooters and shared bicycles) and ride-hailing services (e.g., UBER), which vary according to location. A map of the study area including the PT network is presented in Fig.~\ref{fig:study_area}.

\begin{figure*}
    \centering
    \includegraphics[width=18cm]{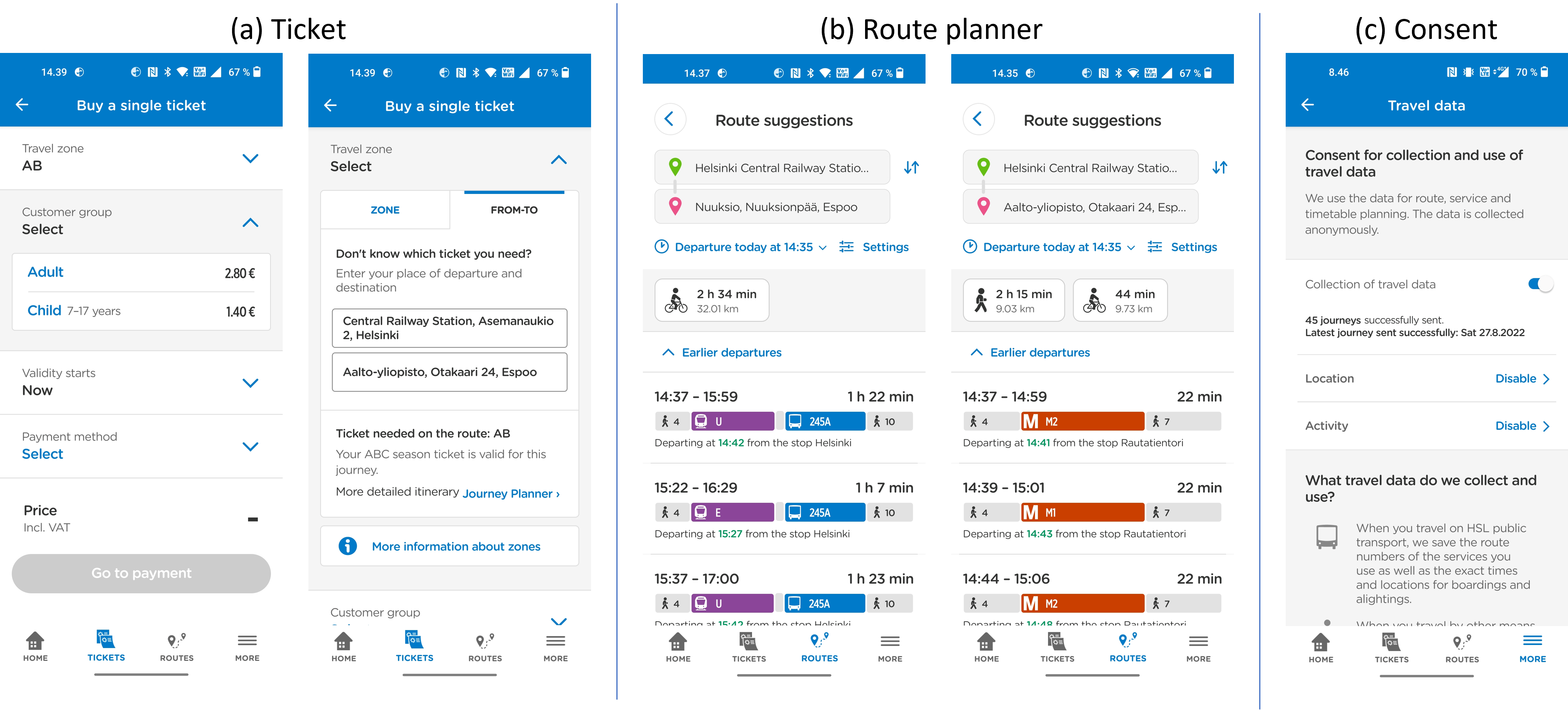}
    \caption{Sample of HSL mobile phone app interface. }
    \label{fig:hsl_app_ui}
\end{figure*}

HSL provides PT services for 9 municipalities (HSL area) which, in addition to the capital region, include Siuntio, Kirkkonummi, Sipoo, Kerava and Tuusula \cite{HSLArea}. To facilitate PT trips HSL offers a mobile phone ticketing application that provides travelers with information on PT operation (e.g., fare zones, real-time PT vehicle locations, timetables, best routes, delays, and PT news, among others). Fig.~\ref{fig:hsl_app_ui} shows a sample of the HSL mobile app interface and features.

\subsection{Data collection infrastructure}
\label{sec:collection_infra}

\begin{figure}[h]
    \centering
    \includegraphics[width=8cm]{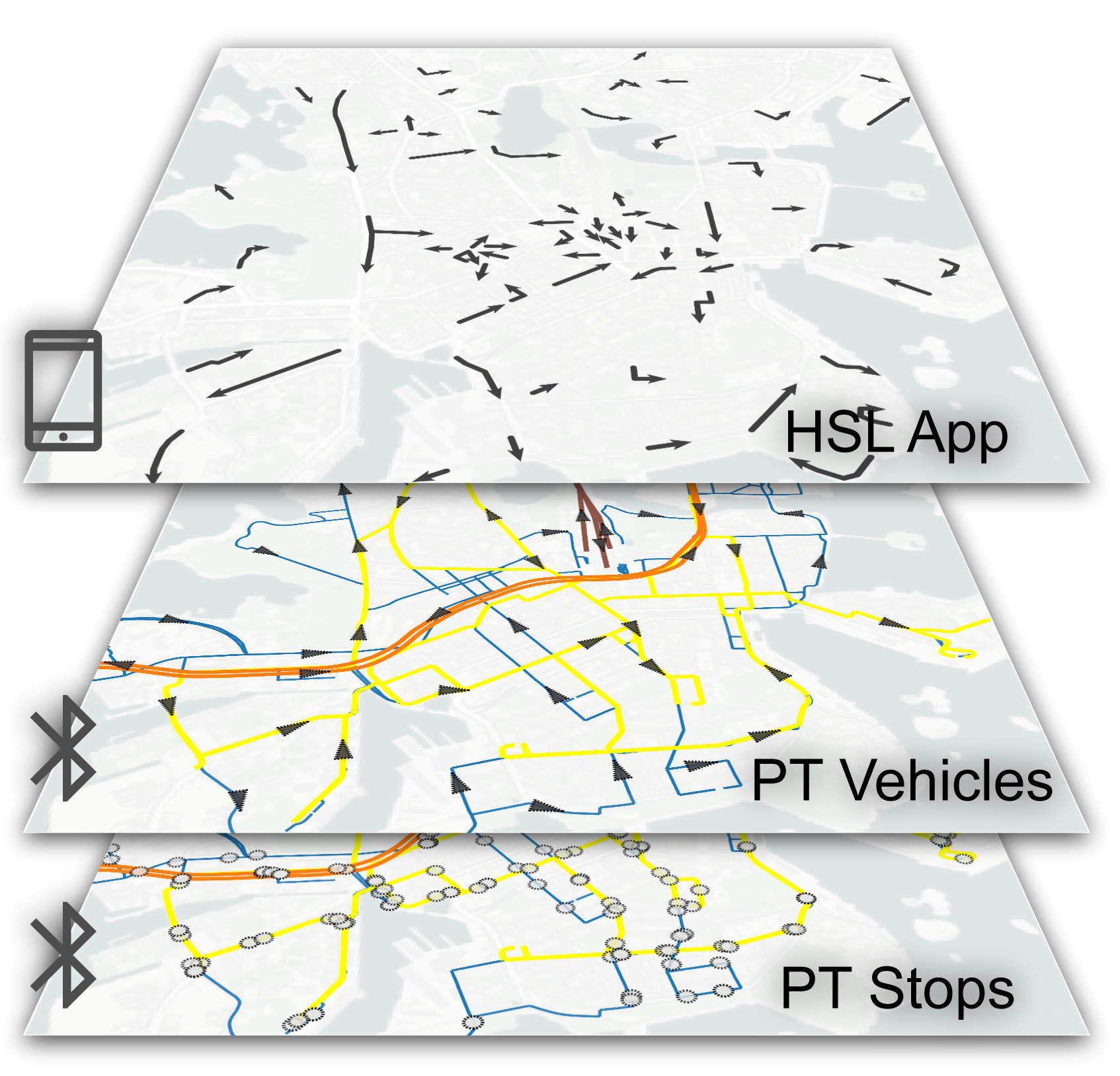}
    \caption{Physical infrastructure of data collection.}
    \label{fig:infrastructure}
\end{figure}

The data collection infrastructure (Fig. \ref{fig:infrastructure}) relies principally on the HSL mobile app. Through the app, a user's device is assigned a random ID for every day that shares data, and is able to recognise Bluetooth low energy beacons throughout the PT network \cite{beacon}.
The app also uses the mobile phone's GPS coordinates to determine the grid cells through which the user moves. Finally, the app uses the activity recognition modules of the user's mobile phone to determine whether the user is still, walking, cycling, or on board a vehicle.

Therefore the physical sources used in the data collection can be classified in the following way:
\begin{itemize}
    \item Stationary Bluetooth beacons. These are installed throughout the study area at PT stops (bus and tram stops, train platforms in stations, and metro stations).
    \item Moving Bluetooth beacons. These are installed inside PT vehicles (buses, trams, trains, metros and ferries).
    \item Portable devices. These are the users' mobile phones, which recognise the user activity and movement and also recognise the Bluetooth beacons throughout the PT network.
\end{itemize}

The HSL app logs beacon recognition and pushes this information along with position and activity information to the TravelSense servers. The TravelSense data collection system is based on the framework described in \cite{Rinne2017}. Currently the data collected is from users who have explicitly opted-in to share their mobility data with HSL (see Fig.\ref{fig:hsl_app_ui}(c)).

\subsection{Available dataset}
\label{sec:data}

In this paper, data corresponding to the time period from the 12th to the 31st of November 2020 are used. The information is structured based on \emph{legs} and \emph{trip chains}, with the following definitions:
\vspace{2mm}
\begin{itemize}
    \item \emph{Leg - Individual segment within a trip chain recognised by the data collecting system and pre-processing as a discrete stage within the journey, either because there are pauses in the movement, or there is a change in recognised activity.}
\vspace{2mm}    
    \item \emph{Trip chain - A series of legs that have been recognised by the system in pre-processing as being part of a single journey. The end-points of trip chains are recognised by prolonged periods of remaining in the same location and no significant changes in activity.}
\end{itemize}
\vspace{2mm}
The dataset captures details of door-to-door journeys whilst preserving enough anonymity to avoid identification of individuals. Each user's mobile phone is assigned a random ID for every day that the user shares data. Within the PT network, the location of individuals is determined by recognition of Bluetooth beacons which are located at bus stops (only within the Helsinki city limits), train and metro stations, as well as inside PT vehicles. For locations outside the PT network GPS data is used, actual coordinates are not recorded, but reported in a coarse grained manner. Locations of app users are resolved up to grid cells of dimension $250\, \textrm{m} \times 250\, \textrm{m}$, into which the region served by HSL has been subdivided. For privacy purposes, the timestamps of sections of journeys outside the PT network are obfuscated by rounding to the nearest quarter hour, either backwards or forwards in time. 

The raw information included in the available dataset differs depending on whether it is derived from outside or inside the PT network. More specifically, for app users outside the PT network it includes:
\vspace{2mm}
\begin{itemize}
\item Timestamps of start and end of journey legs (rounded to nearest quarter-hour)
\item Grid cells associated to each journey leg
\item Type of movement (walking, cycling, or vehicle),
\end{itemize}
\vspace{2mm}
\noindent while for legs of journeys within the PT network it includes:
\vspace{2mm}
\begin{itemize}
    \item Time stamps of start and end of leg
    \item PT stop IDs and coordinates where leg starts and ends
    \item PT mode used
    \item PT line used (and direction)
\end{itemize}
\vspace{2mm}

\begin{figure}[]
    \centering
    \includegraphics[width=8cm]{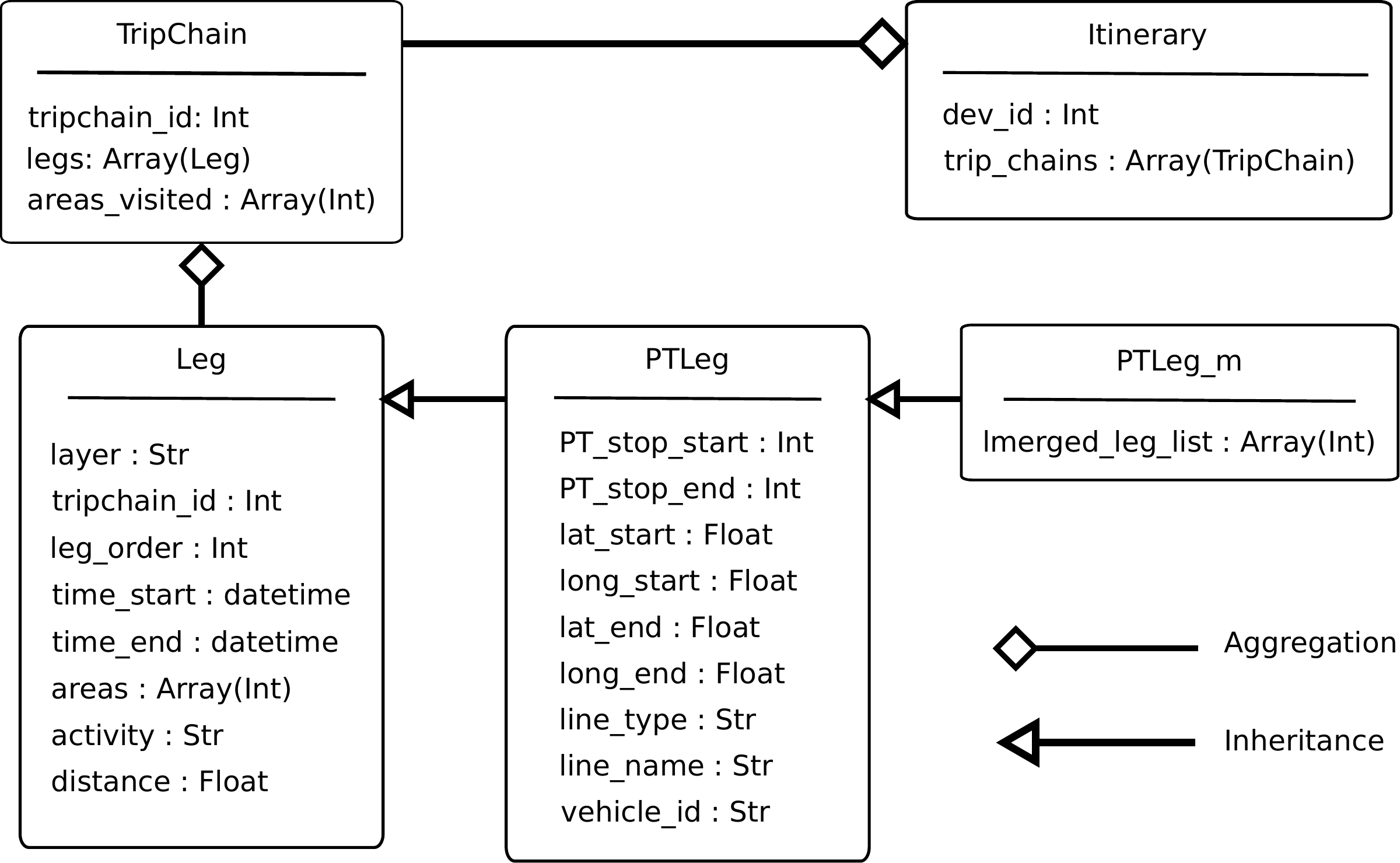}
    \caption{Class diagram for trip objects}
    \label{fig:class_diagram_trip_objects}
\end{figure}



The TravelSense dataset is stored in a database across several tables which need to be collated and then converted into a usable format in order to have all relevant information for a device's itinerary.

In this study, available data is processed using the Python programming language in order to make the handling of the data easier, by using the object-oriented paradigm to have data structures that mirror the \emph{leg} and \emph{trip chain} concepts (defined above) and used by HSL. Classes are defined for \emph{legs} and \emph{trip chains} and an additional \emph{itinerary} class is instantiated for each device ID. The class hierarchy is logically organised so that each itinerary object contains all the trip chains which in turn contain the journey legs associated to them. The relationships between these classes, and some object details are shown in a simplified class diagram in Fig.~\ref{fig:class_diagram_trip_objects}.

It has been noted by Wang and Chen \cite{Wang2018} that among the many transportation studies using mobile phone data, only a few report properties and issues with the data. When using proprietary data is common for there to be restrictions on the properties of the data that can be shared. However, as it is noted by the above study, the processing can determine whether biases or inaccuracies in estimations are introduced. Thus, we summarise some of the issues that have been identified in the data:

\begin{enumerate}
    \item Spurious legs when travelling in PT modes. Erroneous legs appear in the dataset when there exist prior an subsequent legs carried out in a specific PT vehicle.
    \item Conjoined trip chains. Legs are identified as part of the same trip chain even when there are significant time and spatial gaps between some of the legs.
    \item Delayed boarding/anticipated alighting. Test journeys by HSL showed that occasionally boarding and alighting stops recorded correspond to later and earlier stops on the routes, respectively, rather than to the actual stops used in the journey. This results in shorter PT legs and additional spurious legs.
\end{enumerate}

Of these issues, the most easy to identify and correct is issue 1, which occurs in around 4\% of trip chains that have at least one PT leg. Spurious legs are identified by checking all trip chains that have more than one leg carried out by PT. Successive PT legs are checked, if they have been undertaken in the same vehicle then these and all other legs occurring in between are merged into a new longer leg (class \verb|PTLeg_m| in Fig.~\ref{fig:class_diagram_trip_objects}).
Issues 2 and 3 are harder to identify and require additional analysis. These fall beyond the scope of this study and will be addressed in future research.

\section{Multi-modal mobility patterns and validation}

\begin{figure*}[h]
\centering
\includegraphics[width=17cm]{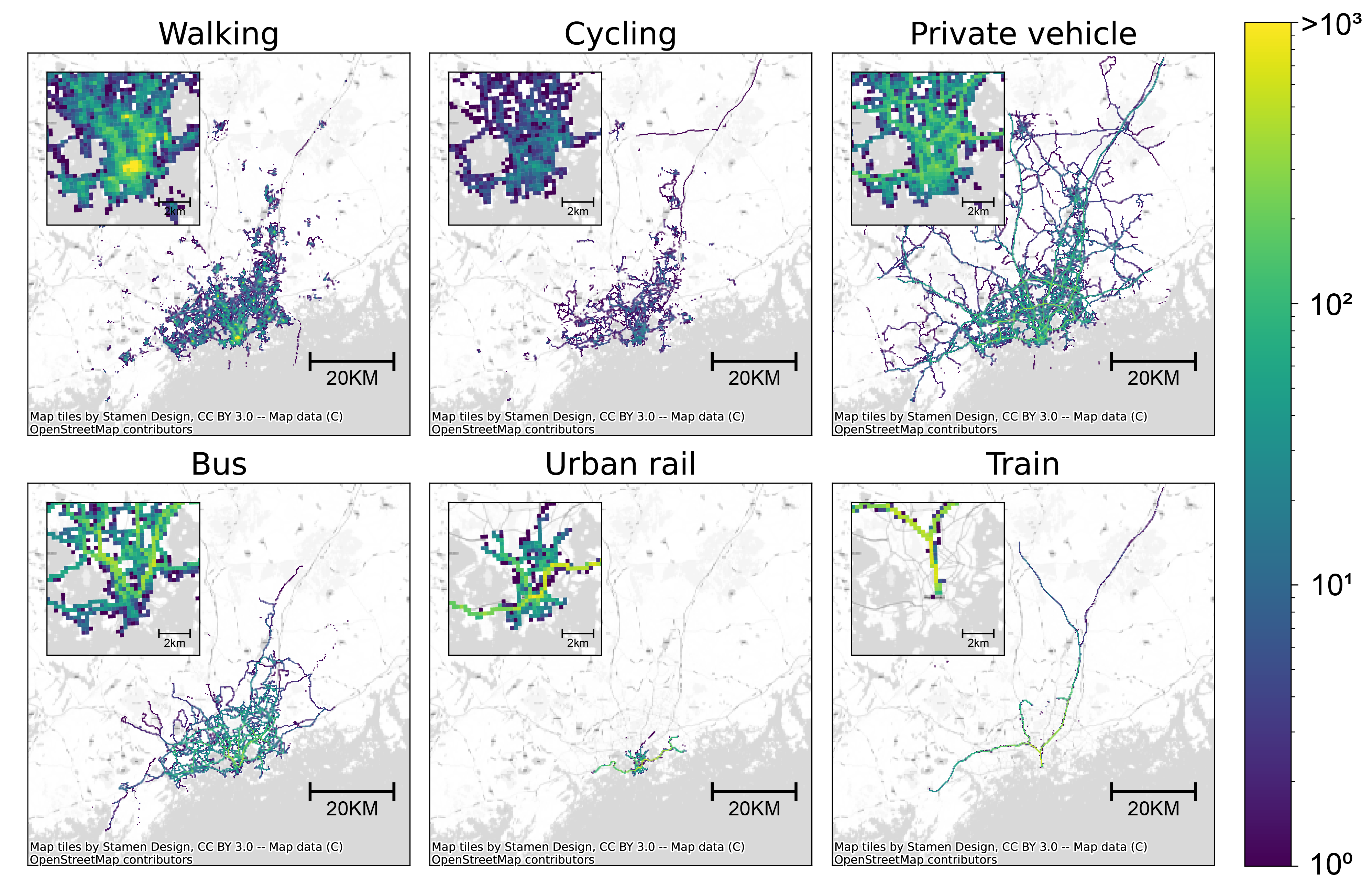}
\caption{Amount of recognised legs per transport mode displayed for each grid cell. The inset shows the center of Helsinki.}
\label{fig:leg_count_heatmap}
\end{figure*}

\begin{figure*}[h]
\centering
\includegraphics[trim = 0mm 150mm 0mm 0mm, clip, width=14cm]{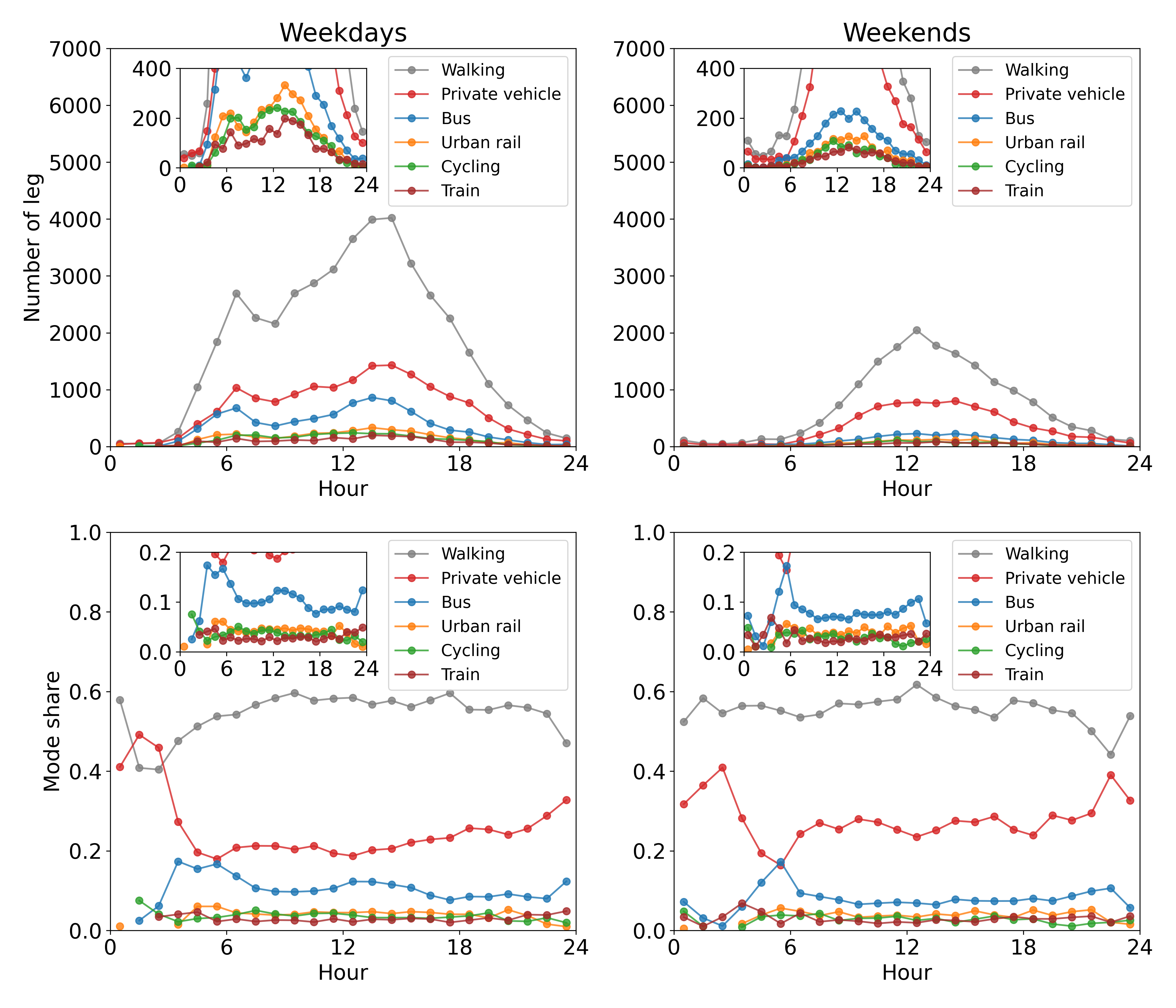}
\caption{Amount of recognised legs per transport mode displayed for each hour. The inset shows the leg number for cycling, urban rail, and train.}
\label{fig:leg_count_hourly}
\end{figure*}

In this study, the focus is on depicting six primary traffic modes, which account for 98\% of total travel demand. Walking, cycling, and private vehicle modes (Private modes) are collected by the HSL mobile app. Bus, urban rail, and train modes (PT modes) are collected by Bluetooth beacons installed in stops and HSL vehicles. Urban rail is comprised of both tram and subway modes. Given the leg is the atomic unit of TravelSense data, the analysis in this session, is performed on legs. 

\subsection{Spatio-temporal distribution of multi-modal mobility patterns}

In order to analyze the spatio-temporal distribution of multi-modal mobility patterns, the first step was to investigate how mobility patterns associated with different modes are distributed in HSL area. A visual representation of travel demand per mode is included in Fig.~\ref{fig:leg_count_heatmap}. Travel demand is expressed by recognized legs within each grid cell. Generally, even at this coarse-grained visualization, the demand patterns trace out the road and PT network (see Fig. \ref{fig:study_area}) as well.

To quantify the coverage for each traffic mode, the number of used (i.e., leg number $\geq$ 1)  grid cells are divided by the total number of grid cells in HSL area. As Table \ref{tab:spatial_distribution} shows, private vehicle is the mode with the largest coverage 39.92\%. Private modes sensed by the HSL mobile app have larger coverage than PT modes sensed by Bluetooth beacons. This result demonstrates the importance of the mobile app, which significantly enhances sensing ability beyond the PT system. Using TravelSense data, almost half of the HSL area could be sensed. It is noted that most sensed grid cells with a high population. 

\begin{table}[b]
\centering
\caption{Spatio-temporal characteristics of TravelSense data}
\label{tab:spatial_distribution}
\begin{tabular}{@{}cccccc@{}}
\toprule
Modes           & Coverage(\%) & Mean   & Std.   & $\alpha$ & $R^2$ \\ \midrule
Walking         & 22.91        & 20.07  & 78.22  & -1.187   & 0.95  \\
Cycling         & 13.44        & 3.65   & 4.73   & -1.009   & 0.66  \\
Private vehicle & 39.92        & 25.40  & 48.33  & -1.277   & 0.96  \\
Bus             & 17.26        & 28.94  & 51.05  & -1.296   & 0.98  \\
Urban rail      & 1.49         & 112.06 & 156.48 & -1.681   & 0.64  \\
Train           & 2.86         & 81.18  & 130.85 & -1.539   & 0.91  \\ \midrule
Total           & 46.21        & 52.40  & 134.82 & -1.368   & 0.97  \\ \bottomrule
\end{tabular}
\end{table}

As shown in Table \ref{tab:spatial_distribution}, urban rail has the largest mean leg number of grid cells. PT modes have higher leg numbers than private modes, which reflects the PT system's role in the urban transportation system for providing large-capacity transportation. Moreover, the standard deviation of leg numbers for each mode is larger than the mean of leg numbers, which implies the heterogeneity of leg number distribution. A power law distribution was used (i.e., $\hat{n}=n^\alpha$, where $n$ is the leg number of each grid cell, $\alpha$ is the power law coefficient) to capture the heterogeneity of leg number distribution. The $R^2$ (i.e., Coefficient of determination) of each mode is larger than 0.6, which shows the leg number distribution can be fitted with a power law distribution. The power law coefficients for urban rail and train are less than $-1.5$, which means larger heterogeneity, some grid cells or route intervals may have high occupancy.

The following step was to investigate the travel demands of different traffic modes in terms of the temporal domain. Due to the commuting patterns in urban areas, legs were separated into weekdays and weekends, and then legs were aggregated into one-hour time bins according to the start time of a leg. Fig.~\ref{fig:leg_count_hourly} shows the hourly number of legs per mode. In most cases, the common trapezoid shape of curves representing demand per hour is observed. During weekdays, all modes have two distinct peak hours. All modes except for cycling reach the morning peak between 7 am and 8 am, while afternoon peaks appear between 1 pm and 4 pm. The afternoon peaks are earlier than in many cities partly because the sunset time in Helsinki is around 3:45 pm during November and, traditionally, the Finnish work day is from 8 am  to 4 pm. The length between two peaks for walking and private vehicles is almost 8 hours. For weekends, all modes show single afternoon peak patterns, the peak timing is around 2 pm. The average Pearson coefficient correlation of temporal travel demands between any two modes during weekdays is 0.957, and 0.962 for weekends, which shows all modes shared similar temporal patterns. 

As TravelSense only collects data from HSL mobile app users who opt-in, the amount of volunteers is sparse compared to the total population living in HSL area (under 1\% of PT ridership for the study period). In addition, it is reasonable to expect a bias towards travelers who are primarily PT users. However, the spatio-temporal patterns show a high degree of consistency with experts' domain knowledge. To quantify how representative the TravelSense data is, it is compared here with two widely used data sources: mobile phone trip data and travel survey data. These types of data are usually treated as near-ground-true data.

\subsection{Comparing with mobile phone trip data}

\begin{figure*}[h]
\centering
\includegraphics[width=17cm]{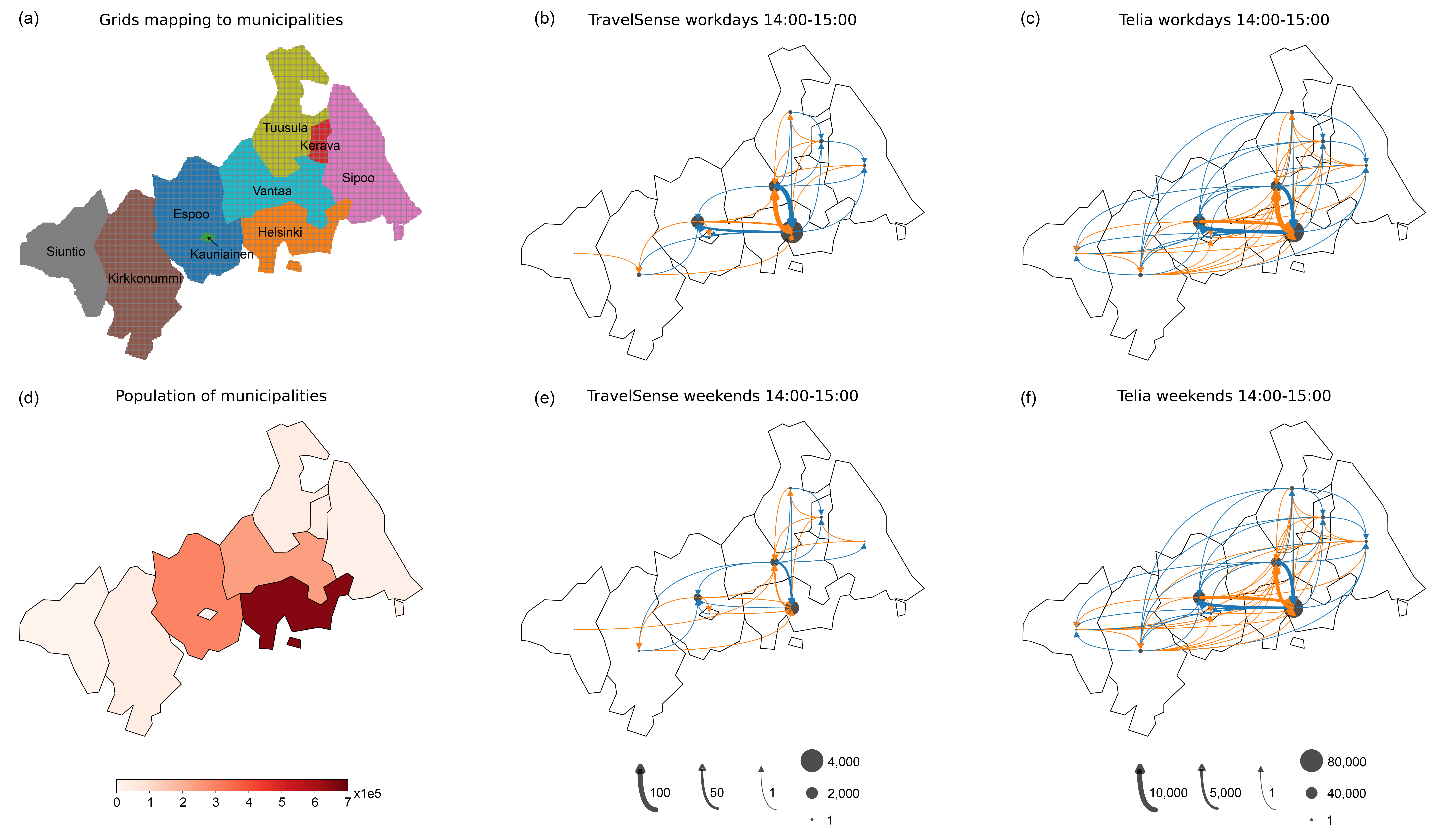}
\caption{Comparison between TravelSense and mobile phone trips data. (a) is the mapping relationship between grids and municipalities; (d) is the population of each municipality; In (b), (c), (e), (f), Curves present the mobility flow between two distinct municipalities, circles present the mobility flow within the same municipalities.}
\label{fig:vs_telia_case}
\end{figure*}

Mobile phone trip data is provided by the mobile network operator Telia, which provides telecommunication services to one-third of the Finnish population. This dataset contains the hourly average OD flow $T^{\textnormal{Telia}}(o, d, h)$ between Finnish municipalities during workdays and weekdays for each month. To protect users' privacy, all OD flows are aggregated and anonymous. Here, we used mobile phone trip data during November 2020, the same period as TravelSense. More details of the mobile phone trip data can be found in a previous study by Kiashemshaki et al. \cite{Kiashemshaki2022}. 

To compare TravelSense data with mobile phone trip data on the same scale, a two-step approach was used:
\begin{enumerate}
    \item We mapped the started grid and ended grid of a leg to municipalities, the mapping relationship between grids and municipalities see Fig.~\ref{fig:vs_telia_case}(a).
    \item We aggregated the trips of TravelSense during 1-hour time window $t$ from the municipality $o$ to the municipality $d$ as $T^{\textnormal{TravelSense}}(o, d, h)$. Workdays and weekends are aggregated separately.
\end{enumerate}

For example, Fig.~\ref{fig:vs_telia_case}(b) shows the 14:00-15:00 OD matrix for workdays from TravelSense, while Fig.~\ref{fig:vs_telia_case}(c) shows it from the mobile phone trip data. The TravelSense data is able to capture mobility flows with large magnitudes, however, some long-distance trips observed in the mobile phone data are missing. There are two possible reasons, the first is the low amount of HSL mobile app volunteering users compared to mobile phone users, which is within an order of magnitude of the whole population level. The second is that current OD flows are estimated by legs, which are shorter than trip chains. Hence, the accuracy could be improved by considering trip chains in future research.

\begin{figure}[h]
\centering
\includegraphics[width=8cm]{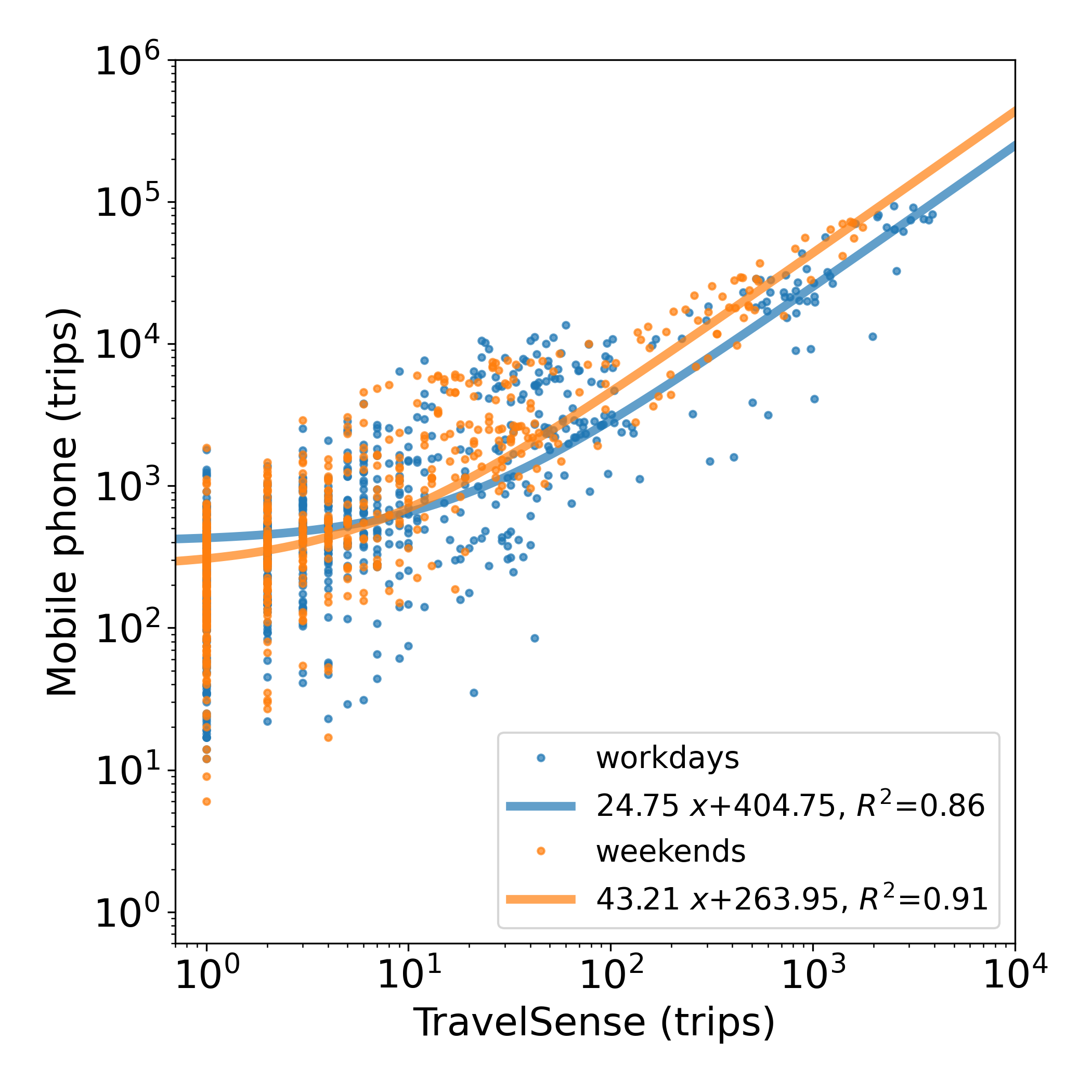}
\caption{Correlation between TravelSense OD flow and Mobile phone OD flow for workdays and weekends. Lines are the fittings.}
\label{fig:vs_telia}
\end{figure}

We performed linear regression analysis to quantify the correlation between TravelSense OD matrices $T^{\textnormal{TravelSense}}(o, d, h)$ and mobile phone record (i.e. Telia) OD matrices, $T^{\textnormal{Telia}}(o, d, h)$:
\begin{equation}
    T^{\textnormal{Telia}}(o, d, h) = \alpha T^{\textnormal{TravelSense}}(o, d, h) + \beta,
\end{equation} where $T^{\textnormal{TravelSense}}(o, d, h)$ presents the leg number between two municipalities $o$ and $d$ during a 1-hour time window $h$, while $T^{\textnormal{Telia}}(o, d, h)$ presents the trip number in mobile phone trip data. Parameter $\alpha$ is the slope, and $\beta$ is the intercept in the linear regression.

As shown in Fig.~\ref{fig:vs_telia}, TravelSense has high positive correlations between mobile phone trip data. The $R^2$ value for workdays is 0.86, while the $R^2$ for weekends is 0.91. The high correlations show the representativeness of TravelSense, which is able to capture the backbone of Helsinki's mobility patterns despite the sparsity of the dataset. There is a potential in future works to scale the TravelSense data up according to its relationship with mobile phone trip data.


\subsection{Comparing with travel surveys}

Although the trip-based OD matrices inferred from mobile phone data are regarded as near-ground-true data, mobile phone trip data usually lack transport mode information. The travel survey contains comprehensive information about transport mode share. However, the large-scale travel survey in HSL servicing is performed every 5 years. We compare the mode share ratio of TravelSense legs with results from existing surveys for years 2012 \cite{TravelHabits2012} and 2018 \cite{TravelHabits2018} for validation purposes.

\begin{table}[b]
\centering
\caption{Mode share comparison of the TravelSense data with HSL travel surveys of 2012 and 2018}
\label{tab:mode_share}
\begin{tabular}{@{}llllll@{}}
\toprule
Percentage(\%)          & Car & PT & Cycling & Walking & Other \\ \midrule
TravelSense legs        & 26  & 19 & 4       & 38      & 13   \\
TravelSense trip chains & 19  & 41 & 4 & 19  & 17     \\
Travel surveys (2018)   & 39  & 22 & 9       & 29      & 1     \\
Travel surveys (2012)   & 40  & 24 & 8       & 25      & 3     \\ \bottomrule
\end{tabular}
\end{table}

Table \ref{tab:mode_share} shows the mode share derived from the HSL mobile app. Modes designated as `other' in TravelSense data are either `running' or not recognized. Walking is the dominant mode based on TravelSense legs, unlike the results of the two surveys that indicate private car as the most used mode. While it is expected that when counting legs rather than whole trips walking can be overrepresented, additional causes could be erroneous split of legs corresponding to walking, the sensitivity of beacons and/or mobile phones, and effects of the COVID-19 pandemic, among others (see Section  \ref{sec:data} for more details). The percentages of travelers using car and cycling are lower than those of the travel surveys. While PT legs have a percentage of 19\%, which is close to the two travel surveys' results. Interestingly, when aggregated into trip chains, the mode share of PT trips jumps to 41\%, which supports the conclusion that the TravelSense data is biased toward PT users.

In this section, we have focused on demand patterns while considering transport modes independently. This is especially important during the validation stage with coarse-grained data. Of special interest to PT practitioners, however, is to understand travel behavior where single trip chains span several modes. We now turn to this in the following section, where we analyze transfers between PT modes.

\section{Applications for public transport: A preliminary transfer analysis}

Data are at the core of methods, models, and tools for better understanding and hence improving transport services. PT is the backbone of urban mobility and lack of data often leads to uncertainties regarding the operation of PT services. Among the wide range of available transport data technologies and sources \cite{Welch2019}, smartcard and GPS tracking are two of the mostly used for reconstructing traveler trajectories as part of understanding traveler behavior and quantifying PT performance. As discussed, these data sources are often problematic for studying transfers, which are critical within a multi-modal trajectory since they cause significant disutility (e.g., by introducing extra walking and waiting times \cite{Kujala2018}). PT operators need more and more accurate information of this part of the traveler trajectory \cite{Nassir2015, Khalil2021} to improve their systems' performance and attractiveness.

\begin{figure}
    \centering
    \includegraphics[width=9cm]{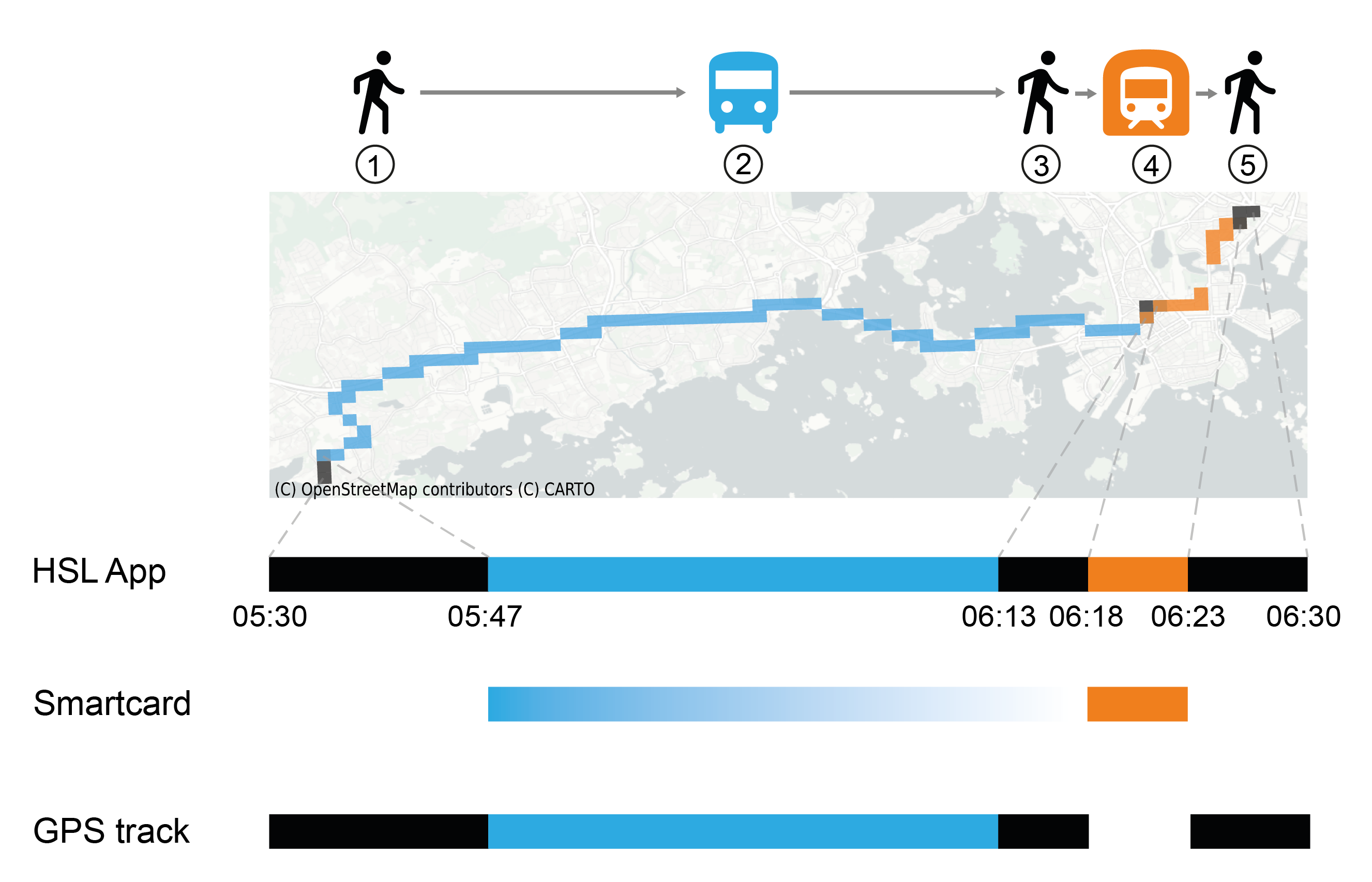}
    \caption{Visualisation of trajectory data and comparison with other data sources.}
    \label{fig:sample_trajectory_visual}
\end{figure}

TravelSense data allow for such detailed representation since the focus is not just on the leg level, but also on the potential of sensing full trip chains. Fig.~\ref{fig:sample_trajectory_visual} presents an example of information on the traveler trajectory from origin to destination available in the dataset. In the exemplar trajectory, a traveler:
\begin{enumerate}
    \item walks from origin to a bus stop (black bar),
    \item experiences travel time on-board the bus (light blue bar)
    \item transfers to a train stop by walking (black bar)
    \item experiences travel time on-board the train (dark blue bar)
    \item walks to the destination (black bar).
\end{enumerate} Fig.~\ref{fig:sample_trajectory_visual} also illustrates the difference between the available TravelSense trajectory data and other sources. Smartcard data cannot record walking legs and, in some cases, can contain only tap-in data (fading light blue) \cite{Zheng2018, Tu2019}. GPS tracks do not record transport mode \cite{Zheng2008a} or underground segments \cite{MSA2011}.

Transfers are usually defined with strict distance and time thresholds between alighting and boarding locations and times (e.g., \cite{Yap2019}). In this preliminary analysis of transfers within the HSL network, a PT transfer is considered to take place any time there are two subsequent PT legs in a single trip chain regardless of any intervening leg(s) on other modes. This is necessary, since walking legs between transfers are recognised in the dataset, but also allows a more flexible view of how travelers use the PT network. 

\begin{figure}[b]
    \centering
    \includegraphics[width=0.45\textwidth]{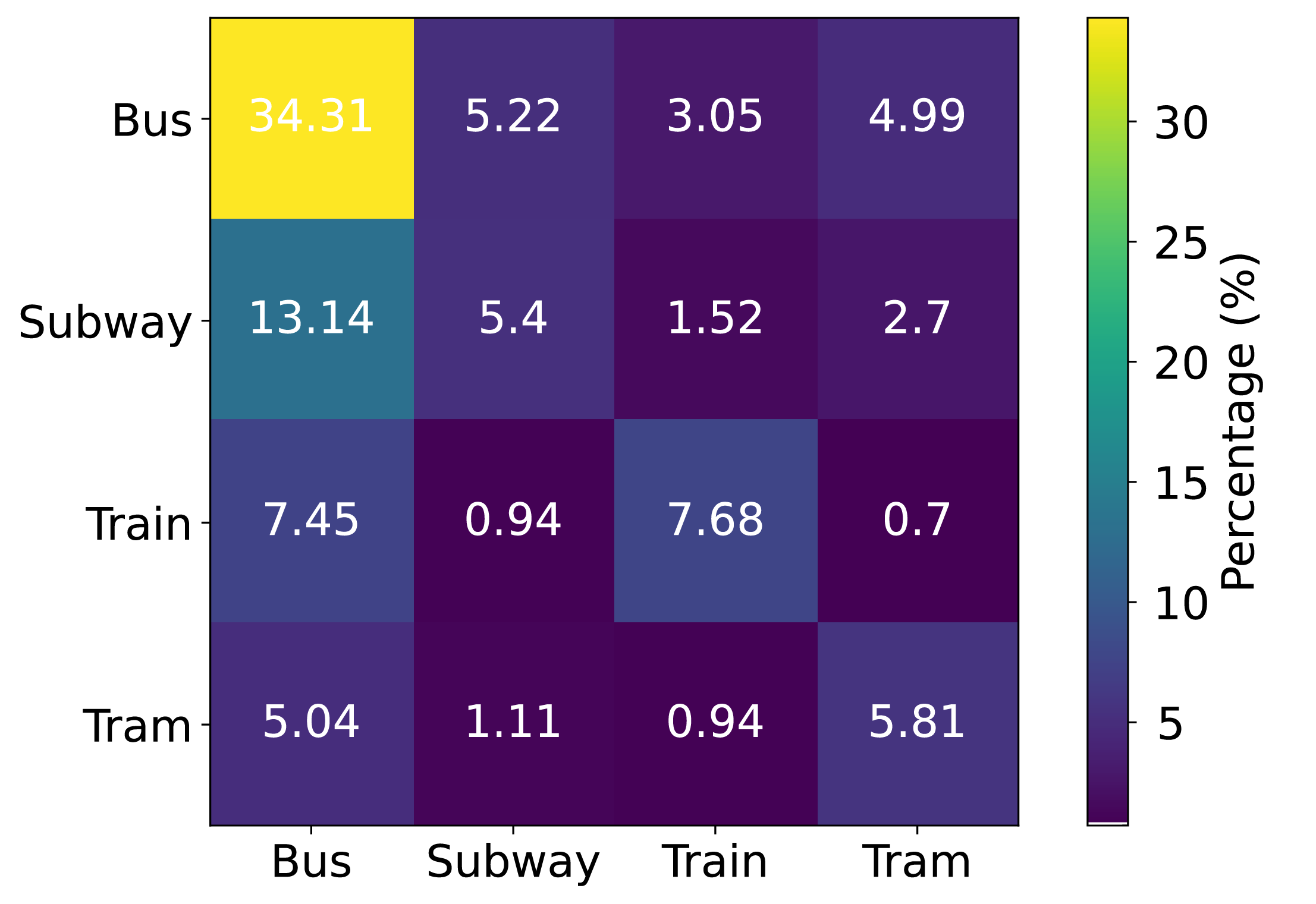}
    \caption{Intermodal PT transfer matrix. Proportions of transfers between PT modes are shown.}
    \label{fig:PT_transfer_matrix}
\end{figure}

Fig.~\ref{fig:PT_transfer_matrix} shows the transfers between PT modes observed. We can see that in Helsinki, transfers involving buses are the most common kind, with bus-to-bus transfers making up more than one-third of all observed transfers in the time period covered by the data. It can also be seen that the bus network plays an important supporting role to the subway and that train-to-train transfers are also significant. While this already yields important insights into multimodal PT journeys, they can still be enhanced further by looking at how the PT infrastructure is used.

PT stops can be clustered to allow more aggregated analyses of the resulting `hubs'. In this study, PT hubs are identified by clustering stops according to geographical proximity using the DBSCAN algorithm \cite{DBSCAN}. The parameters used were $\epsilon=1.255\times10^{-5}$ radians (equivalent to about 80m), and the minimum points in a cluster core were $M=4$. Parameter $M$ is also the minimum number of stops in a cluster and was chosen as 4 to have at least as many stops in a cluster as would be needed to include the stops for two PT routes (e.g., bus) in both directions of travel. Parameter $\epsilon$ was chosen initially using the elbow method and refined by inspection of the largest clusters, such as the main train station, so that PT stops clearly designed to be part of the hub were included.

\begin{figure*}
    \centering
    \includegraphics[width=0.8\textwidth]{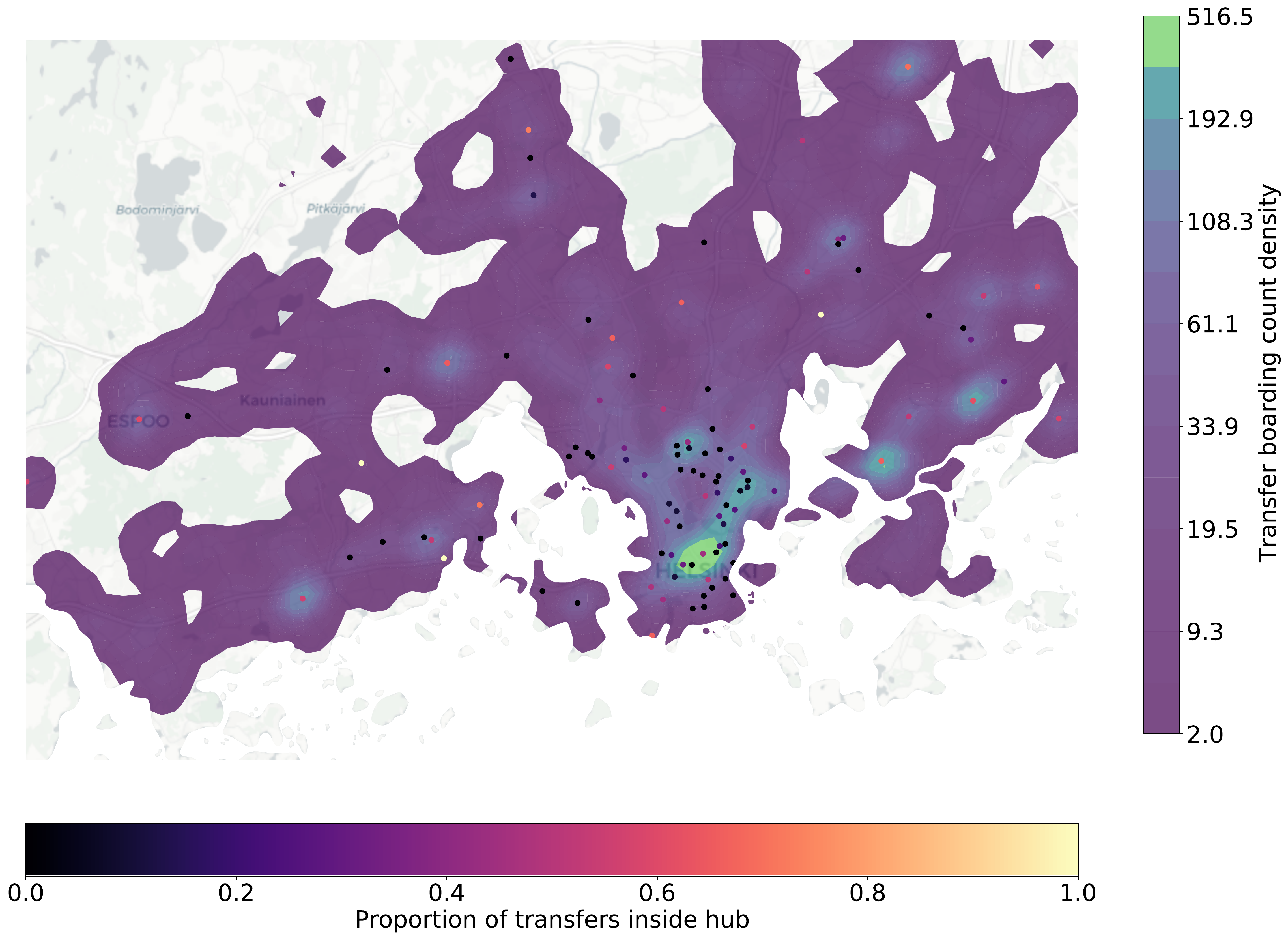}
    \caption{Illustration of spatial distribution of transfer events (heatmap) and proportion of transfers within transfer hubs (colored dots).}
    \label{fig:kde_and_clusters}
\end{figure*}

To understand how hubs are used as parts of traveler trajectories, alighting and boarding stops used for transferring were first identified. Then, for each hub (cluster), the proportion of transfers happening completely within the hub (both alighting and boarding stages) were determined. Fig.~\ref{fig:kde_and_clusters} shows the spatial distribution of transfers in Helsinki with a heatmap showing the kernel density estimate (KDE) distribution of transfer boardings (corresponding colorbar on the right). In addition, the hubs are shown with their colour corresponding to the proportion of transfers happening within them.

As expected, the center of Helsinki has the highest transfer density. Other peaks in density occur in the surrounding areas, mainly around subway and train stations. It is interesting that in the areas outside the central region, the hubs located there tend to have a higher proportion of intra-hub transfers than in the central region. This highlights the need for careful attention on behalf of operators to transfer patterns, if their aim is to ensure a smooth transfer experience for their users. The TravelSense data allow for evaluation of whether hubs are simply of set of geographically related PT stops or whether the PT stops are indeed related via transferring passengers.

The preliminary transfer analysis presented here is a small sample of the opportunities offered by this dataset for PT and transport systems, in general. Comprehensive traveler trajectories allow detailed insights on the real usage pattern of different transport modes, which can be further translated into traveler choices and behaviors, comparison and evaluation of transport modes, quantification and update of well-established performance indicators, among many others. On-going research on this dataset by this group is focused on the identification of hubs with high number of transfers and the analysis of transfer-related hub characteristics.  

\section{Conclusion}

This study focuses on a new mobility dataset, TravelSense, with great potentials for the transport sector, if properly processed and utilized. The dataset is owned and operated by HSL, the Helsinki region transport authority in Finland. Issues encountered while processing the data have been identified and presented here, and the dataset has been validated with external data sources to highlight its promise as a useful resource for sensing and understanding urban mobility. As more Helsinki PT users opt-in to share their data, a comparison with mobile phone trip data suggests that it can attain a high level of accuracy with data from substantially less individuals. This dataset's potential to improve PT has also been demonstrated through a preliminary transfer analysis to enhance and support multimodality. Overall, the TravelSense data can capture richer trajectories than other data sources and can yield actionable insights into urban mobility patterns and usage of PT networks.

\section*{Acknowledgment}
The authors thank HSL for access to the TravelSense dataset, and for their time and discussions about this study. We especially thank Pekka Räty and Tri Quach. Zhiren Huang is supported by the NetResilience consortium funded by the Strategic Research Council at the Academy of Finland (grant numbers 345188 and 345183). Calculations were performed using computer resources within the Aalto University School of Science “Science-IT” project.


\end{document}